\documentclass[11pt,a4paper]{article}
\usepackage[margin=29mm]{geometry}
\usepackage[T1]{fontenc}
\usepackage[utf8]{inputenc}
\usepackage{lmodern,microtype}
\usepackage{amsmath,amssymb,amsthm,mathtools}

\usepackage[authoryear,round]{natbib}
\usepackage[colorlinks=true,linkcolor=black,citecolor=black,urlcolor=blue!55!black]{hyperref}
\usepackage{xcolor}
\usepackage{setspace}
\numberwithin{equation}{section}
\newtheorem{theorem}{Theorem}
\newtheorem{proposition}{Proposition}

\theoremstyle{remark}

\newcommand{\E}{\mathbb E}
\newcommand{\Var}{\operatorname{Var}}

\newcommand{\R}{\mathbb R}
\newcommand{\W}{\mathcal W}
\newcommand{\calF}{\mathcal F}
\title{Optimal Regulation and Information Institutions\\with Hidden Network Equilibria}
\author{Christian Jaag\thanks{Swiss Economics, Ottikerstrasse 7, 8006 Z\"urich, Switzerland and Somaha Foundation, Weinbergstrasse 102, 8006 Z\"urich, Switzerland. Corresponding author: \texttt{christian.jaag@somaha-stiftung.ch}.}}
\date{12 October 2026}
\hypersetup{pdftitle={Optimal Regulation and Information Institutions with Hidden Network Equilibria},pdfauthor={Christian Jaag}}
\begin{document}
\maketitle
\begin{abstract}
Aggregate outcomes can conceal network configurations that expose households to uninsurable risk and require different public services. We study a regulator's joint choice of a corrective activity charge and an information institution in a two-population network game. Although aggregate behavior follows an exact stable law, distinct specialization equilibria require different targeted interventions. With purchased Gaussian monitoring, primitive conditions imply a discontinuous switch between strong regulation without purchased information and informed specialization. The regulator may alternatively obtain signed certification or elicit a privately known state through a report backed by committed random audits. We characterize the three institutions' optimized values and the cost thresholds at which credible disclosure replaces government monitoring. The institutional choice also changes optimal regulation: even when certification and truthful reporting reveal the same state, costly audits favor less specialization and a higher corrective charge. This ordering and monotonicity in audit costs hold under an explicit interiority condition without global concavity of the audit problem. The institutional comparison is solved in the benchmark without feedback from anticipated policy, while the policy switch and strict institutional rankings admit local persistence under additional regularity conditions. Results depend on enforceable verification, restrictions on data access, and an explicitly limited institutional menu.
\end{abstract}

\section{Introduction}

Regulators often observe an aggregate outcome more precisely than the configuration of activities that produces it. A national output measure, sectoral activity index, or average adoption rate may remain stable while firms and households reorganize along different technologies or production links. Those differences matter for public policy when decentralized activity exposes third parties to uninsured risk and when public services deliver different benefits across the resulting configurations. The regulator then faces two distinct instruments: an ex ante charge that discourages risky divergence and ex post targeting informed by costly monitoring.

For example, a confidential standards assessor might verify penalties for technological incompatibilities among interconnected producers, while a separate public agency allocates adjustment capacity to the cluster that needs it most. A verified count of incompatibilities need not disclose which named cluster follows which technology. The information problem becomes consequential when producers' choices also expose households to risks that cannot be insured through verifiable state-contingent contracts. This example motivates a feasible institutional separation; its empirical prevalence and the costs of maintaining it are outside the model.

A conventional regulatory calculation starts with the marginal external damage from an activity and asks how strongly the government should discourage it. This paper adds a policy-design margin. Regulation changes the size of the economic differences about which information might later be acquired. Better monitoring in turn increases the expected benefit of allowing those differences, since the government can better direct a subsequent intervention. When agents foresee this response, information policy also affects their earlier private activity. Consequently, the desired degree of regulation and the value of public monitoring are jointly endogenous.

We develop these effects in a network game with a fixed, two-block interaction structure. Uninsured households bear a risk externality generated by the private activity difference. The government selects a charge on connected agents' activity differences and a precision level for a private Gaussian monitoring signal. After private choices have been made and a stable specialization configuration has emerged, the regulator observes its signal and chooses a contingent public service. The service is then sequentially optimal, and private actors rationally anticipate its expected return.

The analysis combines three results. First, the average activity obeys an exact, stable aggregate equation even when the network supports distinct specialization equilibria requiring different targeted interventions. Second, the expectation of more precisely targeted public services strengthens private specialization under a transparent feedback condition. Third, explicit inequalities in the primitives of network interaction, monitoring technology, regulatory cost, and outside risk guarantee a discontinuous switch between no monitoring with strong regulation and positive monitoring with more specialization. A comparative-static result establishes how uninsured risk, information costs, and targeting benefits shift the critical network exposure. These results establish existence and local robustness for a specified class of networks; they do not establish the same conclusions for arbitrary network topology.

The economic point is a public-finance distinction between instruments that restrain an externality before its realization and institutions that reveal its subsequent allocation-relevant configuration. An institutional-choice result places purchased governmental monitoring alongside signed certification and incentive-compatible reporting with committed random audits. The cost thresholds identify when private information displaces government monitoring. A further regulatory result shows that even equally informative institutions lead to different ex ante corrective charges: the expected audit expenditure needed to deter false reports increases with the permitted specialization amplitude. Higher unit audit costs therefore weakly raise the optimal charge within the active reporting institution. These results require robust reporting incentives and do not treat a favorable cheap-talk equilibrium as guaranteed implementation. The exact aggregate law makes the public-finance trade-off transparent: forecasting remains valid while policy-relevant information depends on institutions. We next relate the argument to the literature, present the economic environment, derive private equilibrium and welfare, and compare the available instruments. Proofs appear in the appendix.

\section{Related Literature}

The closest public-economics contributions analyze information and corrective intervention in environments with externalities. \citet{lee2026} establishes that the sign of optimal research-and-development policy can change with spillover-network connectivity in an oligopoly with exogenous links. Their result concerns the balance of local knowledge spillovers and product-market rivalry; our question concerns corrective regulation of uninsured risk and the endogenous institution for learning the realized specialization state. \citet{li_xiao2023} show that taxes and subsidies affect information disclosure in a persuasion model with spillovers; their information originates with a strategic sender. \citet{elnaboulsi2023} study environmental taxation with heterogeneously precise signals and firms' information sharing. \citet{amir2023} investigate first-best intervention in vaccine markets with network and health externalities, while \citet{chakrabarti2025} analyze equilibrium and welfare in social-interaction games with externalities. Our regulator privately purchases precision about the \emph{endogenous realized configuration} and uses that knowledge for a later welfare-maximizing service; the same regulator also chooses an earlier activity-difference charge. This leads to a monitoring-versus-regulation margin that is absent from the benchmark versions of these studies.

The analysis also builds on the economics of coordination and information. The distinction between unverifiable private reports and information purchased by the government follows the incentive-compatibility logic of \citet{crawford_sobel1982}, although we impose an explicit conflict over signed service allocation rather than solving a general cheap-talk mechanism. With multiple informed agents, debate and cross-reporting may improve revelation \citep{krishna_morgan2001,battaglini2002}; payments depending on agreement may also sustain truthful elicitation equilibria \citep{miller2005}. Costly state verification is a central mechanism-design problem beginning with \citet{townsend1979}; \citet{border_sobel1987} characterize randomized auditing, while \citet{mookherjee_png1989} examine auditing alongside redistribution and insurance, and \citet{benporath2014} analyze allocation with costly verification. These papers motivate an explicit comparison of verified reports, signed certification, and purchased monitoring. Our contribution concerns their endogenous choice jointly with regulation of a network-generated household externality, for a deliberately restricted institutional menu. In this menu, unlike in a comparison holding the activity level fixed, the cost of ensuring truthful reports feeds back into the ex ante optimal corrective charge. \citet{cao2025} analyze privacy-preserving computation and auditing in financial reporting; their study supports the feasibility of separating verified statistics from underlying confidential records, while also underscoring the institutional commitment and verification-access assumptions required here. \citet{angeletos2006} study anticipated public policy and equilibrium multiplicity; \citet{angeletos_pavan2007} characterize the welfare and equilibrium use of information in games with strategic interactions. \citet{hellwig_veldkamp2009} show how endogenous information acquisition and coordination can reinforce each other. \citet{james2011} and \citet{kohlhas2020} examine the welfare value of information under active intervention, and \citet{arato2021} investigate policy rules for endogenous information acquisition and unstable informational equilibria. Those papers establish that information and policy interact. Our narrower result adds an exact aggregation counterexample and a primitive sufficient condition for a switch in the joint choice of a corrective charge and private governmental monitoring.

Finally, \citet{galeotti2020} characterize interventions in network games, while \citet{sun2023} and \citet{kor2025} analyze interventions involving network links and marginal incentives. \citet{myatt_wallace2019} and \citet{leister2020} study information acquisition and its use in network games. The present policy instrument is a charge on activity differences on a fixed network, and the information is acquired by the regulator for a later public intervention. In particular, the regulator does not redesign the network links or coordinate private acquisition decisions. \citet{chadha2020} study aggregate play and welfare on networks. The use of the lowest network eigenvalue to characterize strategic-interaction equilibria is established by \citet{bramoulle2014}; our spectral calculation is a transparent specialization rather than a novel network-equilibrium theorem. More recent research, including \citet{murayama2026} on network heterogeneity and information welfare, and \citet{tian2026} on endogenous information and dynamic feedback, further narrows the scope for broad novelty claims. We focus on a specific public-finance mechanism with uninsured outside households, anticipatory private responses, and an endogenous shift between corrective regulation and informationally guided public provision.

\section{Network Economy, Uninsured Risk, and Public Instruments}

\subsection{Private activity and the interaction network}

The economy has two equally sized populations indexed by $g\in\{+,-\}$. Each population consists of a continuum of atomistic agents. An agent chooses an unrestricted real-valued activity $x_i$. A share $1-q$ of its interaction weight falls on the same population and a share $q$ falls on the other population, with $q\in(0,1]$. Within each block, interaction weight is uniform. The symmetric and row-stochastic interaction operator reduces, for block-constant profiles, to
\begin{equation}
W(q)=\begin{pmatrix}1-q&q\\q&1-q\end{pmatrix},
\qquad \lambda_m=1,\quad \lambda_d=1-2q.
\label{eq:W}
\end{equation}
The parameter $\alpha>0$ measures strategic substitution. A regulator levies a nonnegative charge $r$ on the squared differences in connected agents' activities. The charge is an incentive instrument applied to a fixed network; the government does not choose the links. Revenue is returned through non-distortionary lump-sum transfers. The charge is administered by a confidential assessment mechanism that releases a verifiable \emph{even} statistic of connected activity differences. Each producer receives a private invoice computed from the relevant pairwise activity differences, while the policymaker receives the verified, sign-invariant assessment statistic and aggregate receipts. In a block-constant equilibrium, each producer's certified tax base is $4qa$ in either branch. In particular, the information revealed to the policymaker is the statistic $T(x)=\int\!\int w_{ij}(x_i-x_j)^2\,di\,dj$ and the associated payments, not the underlying labeled inputs. A group-label swap leaves both releases unchanged, so $I(S;T\mid a)=0$ for the symmetric prior. Verification without revealing the private inputs is technologically feasible through privacy-preserving network analytics, although its use here is an \emph{assumption about the disclosure architecture}, not a claim that all tax authorities face this constraint \citep{hastings2023,cao2025}. In particular, inexpensive access to labeled records would remove the model's information scarcity.

After private activities are chosen, the government supplies a signed targeted service $t\in\R$. The sign represents a deviation in service allocation from a common baseline, with positive values favoring the first population and negative values favoring the second. An agent of population $g$ receives a direct return $\omega s_g x_i t+\nu s_g t$, where $s_+=1$, $s_-=-1$, $\omega\geq0$, and $\nu\geq0$. The second component captures a redistributive preference over which group receives the service; it does not depend on the individual activity.

An agent's payoff, omitting the lump-sum rebate and a constant, is
\begin{equation}
\begin{split}
u_i(x_i,x_{-i},t;r)={}&-\tfrac12x_i^2
-\alpha x_i\int w_{ij}x_j\,dj
-\tfrac1{32}\int w_{ij}(x_i-x_j)^4\,dj\\
&-\tfrac r4\int w_{ij}(x_i-x_j)^2\,dj
+\omega s_gx_it+\nu s_gt.
\end{split}
\label{eq:utility}
\end{equation}
An atomistic deviation does not change network aggregates or the posterior policy rule. Consequently,
\begin{equation}
\frac{\partial^2 u_i}{\partial x_i^2}
=-1-\frac38\int w_{ij}(x_i-x_j)^2\,dj-\frac r2<0.
\label{eq:concavity}
\end{equation}
Each private best response is uniquely defined. An equilibrium is block-constant because all agents in the same population face the same strictly concave optimization problem and hence choose the same activity. The distributional term $\nu s_g t$ affects preferences over the final allocation without changing the marginal return to private activity. It cancels when equally sized groups are aggregated, leaving the welfare and equilibrium equations below unchanged. Denote these activities by $x_+$ and $x_-$. Their mean and half-difference are
\begin{equation}
m=\tfrac12(x_++x_-),\qquad d=\tfrac12(x_+-x_-).
\label{eq:md}
\end{equation}

\subsection{The uninsured-income-risk externality}

Two households, each exposed to one population's specialization, have consumption $c_g=\bar c+\ell d\xi_g$. The variables $\xi_g$ are independent normal shocks with mean zero and variance $\sigma^2$. The exposure parameter $\ell$ is positive. Individual shocks are noncontractible, and contingent insurance against these shocks is unavailable. The production agents neither bear nor internalize these households' risk.

The households have CARA utility $U(c)=-\exp(-\gamma c)/\gamma$, where $\gamma>0$. Their aggregate certainty-equivalent loss relative to risk-free consumption is
\begin{equation}
\eta d^2,\qquad \eta=\gamma\ell^2\sigma^2>0.
\label{eq:eta}
\end{equation}
We adopt a money-metric social welfare criterion: household certainty equivalents in a common numeraire are added to producers' quasi-linear payoffs. This criterion is an explicit normative assumption, rather than the unweighted sum of CARA expected utilities. The risk exposure of these households is an external effect of the producers' choices. Lack of contingent insurance prevents complete private smoothing of their consumption risks. Since the shocks are independent, verifiable contingent transfers could diversify the exposure. The friction is therefore the absence of any publicly verifiable record of the household-specific shock at the date when a contingent transfer would be due. Under the assumed technology, the government may redistribute a deterministic transfer but cannot condition an insurance payment on $\xi_g$. Such transfers alter expected consumption without changing $\Var(c_g)$; under the money-metric criterion they cannot eliminate the loss $\eta d^2$. If household income shocks were contractible or if a verifiable ex post claims system were available, direct insurance would become a competing policy and the results below would have to be recomputed. The subsequent public service targets production specialization and is not a household-shock insurance contract.

\subsection{Timing, information, and continuation policy}

The government first chooses $r\geq0$ and an information institution. Its options are Gaussian monitoring with precision $\tau\geq0$, perfect signed certification at a fixed real cost, or a direct report supported by a legally committed random signed audit and a capped sanction. Conditional on the chosen institution, it commits to the specified reporting and verification rule; the subsequent public-service decision remains Bayesian-optimal. The Gaussian regime is the baseline analyzed in the next sections, while the joint institutional comparison is made in Section~\ref{sec:disclosure}. The institutional menu excludes hybrid signals and information-release mechanisms whose informative outcomes are not robustly implemented. Before private choices, a fair producer-private coordination signal $S\in\{-1,+1\}$ is drawn and observed by the producing populations, while it is unavailable to the government. Agents coordinate on the corresponding block-homogeneous equilibrium. This is an explicitly selected symmetric sunspot equilibrium, with $\Pr(S=1)=\Pr(S=-1)=1/2$; its selection is not claimed to arise endogenously from the subsequent gradient adjustment. The confidential charge assessment described above remains invariant to the sunspot branch. The government observes its verified output and the aggregate activity but receives no labeled individual records. The public prior conditional on the selected equilibrium amplitude is common knowledge. The producer-private sunspot is a correlation device; its realization, rather than a claim about unmodeled expectation learning, specifies the equilibrium-selection rule. Individual atomistic deviations leave this prior and the continuation policy rule unchanged. Private agents choose activities, anticipating the government's later best response. The government observes $Y=d+\xi$, with $\xi\sim N(0,1/\tau)$ independent of other shocks, and chooses $t$. At $\tau=0$ the observation is uninformative. The government pays real information costs $C(\tau)=c_1\tau+c_2\tau^2/2$, where $c_1,c_2>0$. Information remains private to the regulator; it is used for the subsequent public-service decision rather than disclosed to the agents. Private messages about the branch are unverifiable, and the benchmark reporting institution neither pays report-contingent rewards nor conducts signed verification. We later relax both restrictions. Producers cannot compel the confidential auditor to release labeled observations. In addition, when $\nu>\omega\sqrt{a_{\max}}$ for a bound $a_{\max}$ on feasible equilibrium amplitudes, agents in the first group prefer a higher $t$ and agents in the second group prefer a lower $t$, independently of the realized branch. Simple costless reports by either group therefore do not satisfy the incentive constraints of a fully revealing cheap-talk equilibrium. The resulting incentive conflict prevents fully revealing direct cheap talk by a single spokesperson or a simultaneous pair under the binary reporting protocol characterized in Proposition~\ref{prop:communication}. It leaves open richer communication, incentive contracts, and authenticated reports, which are evaluated explicitly in Section~\ref{sec:disclosure}.

In social-welfare units corresponding to the sum of two representative group payoffs, the direct ex post net surplus from $t$ is
\begin{equation}
\beta dt-\tfrac\chi2t^2,
\qquad \beta=2\omega+\zeta>0,\quad \chi>0,\quad \zeta\geq0.
\label{eq:latepolicy}
\end{equation}
The term $2\omega dt$ adds the activity-dependent private returns, while $\zeta dt$ allows a return accruing to downstream beneficiaries. The pure distributional returns $\nu s_g t$ cancel between the equal-sized groups and thus do not enter aggregate welfare. The sign of $t$ describes the reallocation of a differential service component around the baseline, while its quadratic term is the additional resource cost of adjustment. Since private activities are sunk when $t$ is chosen, sequentially rational ex post policy is
\begin{equation}
t^*(Y)=\frac\beta\chi\E[d\mid Y].
\label{eq:texpost}
\end{equation}
Government purchases and structural-charge rebates are financed by lump-sum revenue. Neither the charge nor its rebate constitutes an additional real resource cost. Administration of structural regulation has a separate real cost $\kappa r^2/2$, with $\kappa>0$.

\section{Decentralized Equilibrium and Welfare}
\subsection{Decentralized equilibrium and sufficient aggregate statistics}

Define $\varepsilon(q)=\alpha(2q-1)-1$. In a symmetric equilibrium with two realized states, let $d=S\sqrt a$ for an amplitude $a>0$. Define the posterior-quality function
\begin{equation}
f(v)=\E_N[\tanh^2(v+\sqrt vN)],\qquad N\sim N(0,1),\quad v\geq0.
\label{eq:f}
\end{equation}
The Bayesian posterior and its relevant average are
\begin{equation}
\E[d\mid Y]=\sqrt a\tanh(\tau\sqrt aY),\qquad
\E[t^*(Y)\mid S]=S\frac\beta\chi\sqrt a\,f(a\tau).
\label{eq:post}
\end{equation}
Introduce $\mu=\omega\beta/\chi\geq0$ and $B=\beta^2/(2\chi)>0$.

\begin{theorem}[Exact aggregate dynamics and hidden equilibria]\label{thm:aggregation}
For every symmetric row-stochastic block network $W(q)$, the gradient adjustment of expected private marginal returns satisfies
\begin{equation}
\dot m=-(1+\alpha)m.
\label{eq:agg}
\end{equation}
This equation is exact for arbitrary individual activity profiles and is independent of $q$, $r$, $\tau$, and $\mu$. For the two block-homogeneous candidate equilibria $d=S\sqrt a$, equilibrium requires
\begin{equation}
qa=\varepsilon(q)-qr+\mu f(a\tau).
\label{eq:amp}
\end{equation}
When $\mu=0$ and $\varepsilon(q)>qr$, the two states $d=\pm\sqrt{[\varepsilon(q)-qr]/q}$ are locally stable under gradient adjustment, whereas $d=0$ is unstable. The mean converges to zero in all cases.
\end{theorem}

The two branches form a symmetric sunspot equilibrium when the producer-private coordination device selects the corresponding branch. Each private producer conditions its strategy on the realized $S$ and, under the common equilibrium amplitude and public policy rule, is best responding to the other producers. The policymaker chooses $r$ and monitoring before $S$ is drawn. The term \emph{stability} refers solely to gradient adjustment with the continuation-policy posterior held fixed at its equilibrium value. The specified sunspot mechanism selects an equilibrium for each signal realization; the gradient adjustment does not itself select a sunspot or model joint expectation learning.

\subsection{How expected targeting changes private incentives}

\begin{theorem}[Anticipation-induced specialization]\label{thm:anticipation}
Let $\varepsilon(q)>qr$ and suppose $q>\mu\tau$. Then equation \eqref{eq:amp} has a unique positive solution $a=a(q,r,\tau,\mu)$. It is locally stable under the fixed-beliefs adjustment specified in Theorem~\ref{thm:aggregation}. At fixed structural policy,
\begin{equation}
\frac{\partial a}{\partial\tau}
=\frac{\mu a f'(a\tau)}{q-\mu\tau f'(a\tau)}>0\qquad(\mu>0,\ \tau\geq0),
\label{eq:dat}
\end{equation}
while
\begin{equation}
\frac{\partial a}{\partial r}
=-\frac q{q-\mu\tau f'(a\tau)}<0.
\label{eq:dar}
\end{equation}
The observed aggregate remains governed by \eqref{eq:agg}.
\end{theorem}

Expected access to better targeted public services raises the private return to specialization. The presence of $\mu$ makes information policy a determinant of the private equilibrium even though the service is provided later. The restriction $q>\mu\tau$ is a sufficient, easily verified weak-feedback condition, not a necessary condition for local regularity.

\subsection{The welfare wedge and the feasible regulatory allocation}

At a block-constant equilibrium, the sum of private production payoffs, before lump-sum rebates and excluding $t$, equals
\begin{equation}
-\left[1+2\alpha(1-2q)\right]a-qa^2-2rqa.
\label{eq:privatew}
\end{equation}
The last term is returned to the agents through lump-sum rebates. Accounting for the uninsured household loss \eqref{eq:eta} and optimizing the continuation policy \eqref{eq:texpost} yields
\begin{equation}
\W(a,r,\tau;q)
=\left[-1-2\alpha(1-2q)-\eta\right]a-qa^2
-\frac\kappa2r^2+Ba f(a\tau)-C(\tau).
\label{eq:welfare}
\end{equation}
Equation \eqref{eq:amp} permits the elimination of $r$. Define
\begin{equation}
E(q)=\frac{\varepsilon(q)}q=2\alpha-\frac{\alpha+1}q,
\qquad r=E(q)-a+\frac\mu qf(a\tau).
\label{eq:rfroma}
\end{equation}
The reduced objective is
\begin{equation}
\begin{split}
\calF(a,\tau;q,\mu)={}&
[-1-2\alpha(1-2q)-\eta]a-qa^2\\
&-\frac\kappa2\left[E(q)-a+\frac\mu q f(a\tau)\right]^2
+Ba f(a\tau)-C(\tau).
\end{split}
\label{eq:reduced}
\end{equation}
Its feasible set satisfies $a\geq0$, $\tau\geq0$, $r\geq0$, and the regularity condition of the selected equilibrium when relevant. The restriction $r\geq0$ must be retained: the government does not have an unrestricted negative structural charge.

\paragraph{Corrective rationale.}
The unpriced risk loss $\eta a$ is borne by outside households. A higher charge $r$ therefore has an externality-correcting role, but it also changes both the amplitude of equilibrium specialization and the value of subsequent monitoring. The government consequently compares an ex ante corrective instrument with an ex post informed allocation instrument, paying real administration and monitoring costs. This trade-off becomes the main object of the subsequent policy analysis.

\section{Optimal Monitoring and Corrective Regulation}
\subsection{A no-monitoring regulatory benchmark}

When the government acquires no information, the posterior expected targeted service is zero and the private anticipation channel vanishes. For cross-group exposure above the private specialization threshold $q_E$, define
\begin{equation}
\begin{gathered}
q_E=\frac{\alpha+1}{2\alpha},\qquad
E(q)=2\alpha-\frac{\alpha+1}{q},\qquad D(q)=2q+\kappa,\\
K(q)=-1-2\alpha(1-2q)+\kappa E(q).
\end{gathered}
\label{eq:primitive_defs}
\end{equation}
The regulator then chooses a remaining specialization amplitude $a$ to maximize the strictly concave function $[K(q)-\eta]a-D(q)a^2/2$ on $[0,E(q)]$. The unique policy is
\begin{equation}
 a^{N}(q,\eta)=\min\left\{E(q),\max\left\{0,\frac{K(q)-\eta}{D(q)}\right\}\right\},
 \qquad r^{N}=E(q)-a^{N}.
 \label{eq:noinfo}
\end{equation}
For an interior solution, $\partial r^{N}/\partial\eta=1/D(q)>0$. The benchmark formalizes the corrective role of the charge: higher unpriced household risk leads to stronger deterrence. Introducing costly monitoring makes the policy choice richer because the state-contingent public service creates a benefit from allowing specialization to occur. The following results determine when this additional margin changes the optimal regime discontinuously.

\subsection{The endogenous threshold for public monitoring}

At $\mu=0$, the feasible amplitude is $0\leq a\leq E(q)$ and policy is $r=E(q)-a$. Define the optimized information surplus
\begin{equation}
\Psi(a)=\sup_{\tau\geq0}\left\{Ba f(a\tau)-c_1\tau-\tfrac{c_2}{2}\tau^2\right\}.
\label{eq:Psi}
\end{equation}
The local information threshold is $a_0=\sqrt{c_1/B}$. A posterior identity gives $0\leq f(v)\leq v$ and $f(v)=v-v^2+O(v^3)$ as $v\downarrow0$. Consequently, $\Psi(a)=0$ for $a\leq a_0$ and $\Psi(a)>0$ for $a>a_0$.

\begin{proposition}[Information-induced nonconvexity]\label{prop:gaussian}
For $a_0>0$ in the feasible interior, the right-hand curvature of the optimized information surplus is
\begin{equation}
\Psi''(a_0+)=\frac{4B^2a_0^2}{2Ba_0^3+c_2}.
\label{eq:curv}
\end{equation}
In particular,
\begin{equation}
\frac{4B^2a_0^2}{2Ba_0^3+c_2}>2q+\kappa
\label{eq:nonconvex}
\end{equation}
ensures a locally nonconvex reduced policy objective after optimizing information, at $\mu=0$. For suitable values of the linear welfare parameter, a common supporting line to the associated nonconvex cost function produces two separated optimal policy choices. This curvature inequality alone neither establishes that both are interior nor establishes a network-induced transition at an admissible value of $q$.
\end{proposition}

The local loss of concavity arises despite strictly convex monitoring costs. The public-policy mechanism is that monitoring becomes valuable only when network interactions generate sufficiently pronounced, welfare-relevant differences among decentralized actors.

\subsection{Existence and comparative statics from public-finance primitives}\label{sec:primitive}

A conditional persistence theorem is presented in Section~\ref{sec:forward}. Here we first derive the existence of a policy switch directly from inequalities in network, monitoring, and welfare primitives. The result requires neither
an assumed pair of global optima nor a nonconvex information-cost technology.

For the benchmark $\mu=0$, recall the definitions of $q_E$, $E(q)$, $D(q)$, and $K(q)$ from \eqref{eq:primitive_defs}. The monitoring entry threshold is $a_0=\sqrt{c_1/B}$.
Whenever $q>q_E$, the feasible amplitude at $\mu=0$ is
$0\leq a\leq E(q)$ and $r=E(q)-a$. At risk-cost parameter $\eta>0$ the
planner's profiled welfare is
\begin{equation}
\overline{\W}(a;q,\eta)
=-\frac{\kappa}{2}E(q)^2+[K(q)-\eta]a
-\frac{D(q)}{2}a^2+\Psi(a).
\label{eq:wprofileprimitive}
\end{equation}
The Gaussian posterior satisfies $f(0)=0$, $f'(0)=1$, and
$0<f(v)\leq v$ for $v>0$. Therefore $\Psi(a)=0$ for
$0\leq a\leq a_0$ and $\Psi(a)>0$ for $a>a_0$.

\begin{theorem}[A primitive condition for a discrete optimal policy switch]
\label{thm:primitive}
Let $\alpha>1$ and $\kappa,B,c_1,c_2>0$. Suppose that some
$q_0\in(q_E,1)$ satisfies
\begin{equation}
0<a_0<E(q_0),\qquad
\frac{4B^2a_0^2}{2Ba_0^3+c_2}>D(q_0),\qquad
K(q_0)>D(q_0)a_0.
\label{eq:primitive_conditions}
\end{equation}
Then there exists a unique $\eta^*>0$ at which the two
monitoring classes are equally attractive and the government's global
optimum at $q=q_0$ and $\mu=0$ contains a no-monitoring choice
$(a_L,0)$ and an informative choice $(a_H,\tau_H)$ with
\begin{equation}
0\leq a_L<a_0<a_H\leq E(q_0),\qquad \tau_H>0.
\end{equation}
Holding $\eta=\eta^*$ fixed, the optimal policy changes discontinuously
from a low-amplitude no-monitoring regime immediately below $q_0$ to an
informative high-amplitude regime immediately above $q_0$.

For all sufficiently small positive anticipation parameters $\mu$, while
holding $B$ constant, at least one such global switch continues to occur
at a nearby network weight $q^*(\mu)$. In both cases the exact stable
aggregate equation $\dot m=-(1+\alpha)m$ remains unchanged.
\end{theorem}

The theorem establishes existence for an open set of primitive parameters.
It does not assert a unique crossing or a unique global optimizer within
each regime. Its construction also makes explicit that $q$ is exogenous
weighted network exposure, while $r$ is an activity-difference charge,
not a direct choice of graph links.

\paragraph{Analytical admissibility check.}
The three restrictions in Theorem~\ref{thm:primitive} are jointly
satisfiable. To demonstrate this without optimizing a calibrated economy,
set $\alpha=3$, $q_0=24/25$, $\kappa=1$, $B=4$, $c_1=1$, and
$c_2=1/10$. Then $q_E=2/3<q_0<1$ and
\begin{equation}
\begin{gathered}
 a_0=\frac12,
 \qquad E(q_0)=\frac{11}{6},
 \qquad D(q_0)=\frac{73}{25},\\
 K(q_0)=\frac{953}{150},\qquad
 \Psi''(a_0+)=\frac{160}{11}.
\end{gathered}
\label{eq:analytic_example}
\end{equation}
Consequently, $0<a_0<E(q_0)$,
$\Psi''(a_0+)>D(q_0)$, and
$K(q_0)>D(q_0)a_0=73/50$.
Theorem~\ref{thm:primitive} therefore guarantees a positive uninsured-risk
parameter and a discontinuous change in the optimal policy at this admissible
network weight. The calculation proves that the set of admissible primitives
is nonempty.

The planner's objective has strictly increasing differences in $(q,a)$
at $\mu=0$ on the feasible domain. Indeed,
\begin{equation}
\frac{\partial^2\overline{\W}}{\partial q\partial a}
=4\alpha-2a+\kappa\frac{\alpha+1}{q^2}
>2\alpha+2>0,
\label{eq:increasingdifferences}
\end{equation}
where the inequality uses $a\leq E(q)\leq\alpha-1$.
Thus increasing cross-population exposure always favors larger
specialization amplitude in the benchmark optimization problem.

\begin{proposition}[Signed threshold effects]\label{prop:signed}
At a benchmark crossing from Theorem~\ref{thm:primitive}, assume that
exactly two relevant globally optimal regimes are interior in $a$,
nondegenerate, and locally unique within their respective classes;
assume $\tau_L=0$, $\tau_H>0$, and $0<r_H=E(q_0)-a_H$.
Write $\Delta a=a_H-a_L>0$ and
\begin{equation}
M=4\alpha-(a_H+a_L)+\kappa\frac{\alpha+1}{q_0^2}>0.
\end{equation}
Then the critical network weight $q^*$ is locally differentiable and,
evaluated at $\mu=0$,
\begin{align}
\frac{\partial q^*}{\partial\eta}&=\frac1M>0,
&\frac{\partial q^*}{\partial c_1}&=
\frac{\tau_H}{\Delta a\,M}>0,\label{eq:sens1}\\
\frac{\partial q^*}{\partial c_2}&=
\frac{\tau_H^2}{2\Delta a\,M}>0,
&\frac{\partial q^*}{\partial B}&=
-\frac{a_Hf(a_H\tau_H)}{\Delta a\,M}<0,\label{eq:sens2}\\
\left.\frac{\partial q^*}{\partial\mu}\right|_{0^+}&=
\frac{\kappa r_Hf(a_H\tau_H)}{q_0\Delta a\,M}>0.
\label{eq:sens3}
\end{align}
The last derivative varies private exposure $\omega$ and the external
beneficiary component $\zeta$ jointly to hold $B=\beta^2/(2\chi)$ fixed.
The derivatives have the same strict signs in a sufficiently small
positive-$\mu$ neighborhood under the stated regularity conditions.
\end{proposition}

Higher uninsured household risk and higher monitoring costs therefore
raise the cross-network exposure required to make informed specialization
optimal. Greater targeting benefits lower this threshold. A modest
strengthening of private anticipation raises it when informationally
induced specialization requires a higher regulatory charge.
With $\eta=\gamma\ell^2\sigma^2$, the results also provide signed
comparative statics for household risk aversion, exposure, and shock
variance. These comparative statics concern an endogenous discontinuous
policy threshold, not a locally linear private equilibrium.

\subsection{Interpretation of the public-finance trade-off}

The welfare expression \eqref{eq:wprofileprimitive} isolates two distinct effects of regulation. A higher charge reduces the amplitude of equilibrium specialization and consequently reduces the uninsured households' expected risk loss. It also reduces the gross return to later targeting and therefore the incentive to acquire a precise private signal. The same network links determine the decentralized equilibrium and the strength of the regulatory trade-off. This makes the appropriate charge dependent on monitoring costs and the technology of public service provision, in addition to the marginal household externality.

The primitive theorem gives an existence result, rather than an unconditional prescription to suppress specialization or to acquire information. At one extreme, strong regulation keeps third-party risk low and renders costly monitoring unnecessary. At the other, specialization generates sufficient targeting benefits to justify positive precision and a lower corrective charge. The endogenous information threshold creates a nonconcavity in the government's objective despite strictly convex information costs. The resulting discrete choice differs from the usual marginal characterization of a Pigouvian tax.

\subsection{Risk and the switching boundary}

The household certainty-equivalent wedge provides an immediate public-finance interpretation of the comparative statics. An increase in risk aversion, exposure to specialization, or uninsured shock variance raises $\eta=\gamma\ell^2\sigma^2$. Proposition~\ref{prop:signed} then implies a higher critical cross-group exposure before informed specialization is preferred. Higher costs of acquiring data have the same qualitative implication, while more productive ex post targeting lowers the exposure threshold. These signed results apply at a nondegenerate two-regime crossing. They do not describe global monotonicity of optimal policies at all possible parameter configurations.

\section{Forward-Looking Policy and Network Exposure}\label{sec:forward}
\subsection{Persistence when future policy is anticipated}

\begin{theorem}[Persistence of the public-policy regime switch]\label{thm:switch}
Fix $\alpha>1$, $\beta,\chi,\kappa,c_1,c_2>0$, and an uninsured-risk parameter $\eta>0$. Suppose that at $\mu=0$ and a network weight $q_0\in(0,1)$, the reduced welfare problem \eqref{eq:reduced} has exactly two separated global optimizers $z_L=(a_L,0)$ and $z_H=(a_H,\tau_H)$ with
\begin{equation*}
0<a_L<a_H<E(q_0),\qquad \tau_H>0.
\end{equation*}
Assume strict second-order optimality in the free directions, a strict one-sided optimum at $\tau=0$ for $z_L$, and a strict welfare gap away from neighborhoods of these points. Suppose, in addition,
\begin{equation}
T_0:=(a_H-a_L)\left[4\alpha-(a_H+a_L)
+\kappa\frac{\alpha+1}{q_0^2}\right]\ne0.
\label{eq:transv}
\end{equation}
Then, for every sufficiently small $\mu>0$ compatible with $\beta=2\omega+\zeta$ and $\zeta\geq0$, there exists a network weight $q^*(\mu)$ close to $q_0$ at which two separated globally optimal policy regimes coexist. On opposite sides of $q^*(\mu)$ their ranking reverses. The changes in structural regulation and signal precision are discontinuous, while aggregate activity obeys the same exact stable equation \eqref{eq:agg} throughout. If $T_0>0$, an increase in $q$ locally favors the more information-intensive regime.
\end{theorem}

The theorem supplies conditional sufficient conditions for a local regime switch. The primitive-conditions existence result in Theorem~\ref{thm:primitive} supplies one class of benchmark economies in which this persistence argument applies, subject to the stated nondegeneracy conditions. The theorem allows other parameters with no switch or with several optimal policies.

\subsection{Robustness to free but noisy administrative information}\label{sec:free_signal}

The benchmark model assumes that the private government signal is obtained only through purchased monitoring. A regulator may also possess inexpensive administrative indicators that convey imperfect information about the direction of specialization. We therefore consider an independently observed baseline signal $Y_0=d+\xi_0$, with $\xi_0\sim N(0,\tau_0^{-1})$ and a fixed precision $\tau_0\geq0$. Purchased monitoring provides $Y_1=d+\xi_1$ with independent noise and freely chosen precision $\tau\geq0$, at cost $C(\tau)$. When $\tau_0=0$ the baseline signal is absent. Both signals are received only after the private activity decisions and before the targeted public service is chosen.

Conditional on amplitude $a$ and a symmetric prior, the combined posterior satisfies
\begin{equation}
 \E[d\mid Y_0,Y_1]=\sqrt a\tanh\!\left[\sqrt a(\tau_0Y_0+\tau Y_1)\right],\qquad
 \E\bigl[\E[d\mid Y_0,Y_1]^2\bigr]=a f\bigl(a(\tau_0+\tau)\bigr).
 \label{eq:free_posterior}
\end{equation}
The government consequently purchases only the \emph{incremental} precision beyond its administrative baseline. For $\mu=0$ the associated gross value of additional monitoring, evaluated relative to a government using $Y_0$ alone, is
\begin{equation}
 \Delta\mathcal I(a,\tau;\tau_0)
 =Ba\left[f\bigl(a(\tau_0+\tau)\bigr)-f(a\tau_0)\right].
 \label{eq:baseline_increment}
\end{equation}
The term in brackets is nonnegative by the Blackwell ordering of the two Gaussian experiments. The total targeted-service gain entering welfare is $Ba f(a(\tau_0+\tau))$, since free information has a real targeting value even when no further monitoring is purchased. When producers anticipate the policy, the positive-amplitude equilibrium condition becomes
\begin{equation}
 qa=\varepsilon(q)-qr+\mu f\bigl(a(\tau_0+\tau)\bigr).
 \label{eq:free_amp}
\end{equation}
Hence the free baseline signal affects both the later targeting benefit and the ex ante private incentive.

\begin{proposition}[Persistence with an imperfect free signal]\label{prop:baseline}
Suppose the benchmark $\tau_0=\mu=0$ has exactly two isolated global policy optima at a network weight $q_0$, satisfying the interior, strict-second-order, strict-separation, and transverse-crossing assumptions of Theorem~\ref{thm:switch}. The low-monitoring optimum has purchased precision $\tau_L=0$, amplitude $a_L<a_0$, and a strict one-sided optimum; the high-monitoring optimum satisfies $\tau_H>0$. Then for every sufficiently small $\tau_0>0$, there exists a switching weight $q^*(\tau_0)$ near $q_0$ at which two globally optimal regimes coexist. One purchases no monitoring, but uses its free signal; the other purchases positive additional precision. The purchased-monitoring decision and corrective charge change discontinuously across this weight. The conclusion persists for sufficiently small positive private anticipation $\mu$ when the same regularity and equilibrium-feasibility conditions continue to hold.
\end{proposition}

The proposition establishes a local robustness property, separate from the primitive global existence theorem. It does not assert persistence at every degree of administrative informativeness. If the baseline observation identifies $S$ perfectly at no cost, the subsequent Bayesian service already uses all branch-relevant information; paid additional monitoring has zero gross value and cannot be optimal at a positive price.

\section{Endogenous Choice of Information Institutions}\label{sec:disclosure}

The preceding analysis optimizes the intensity of a corrective charge and
purchased Gaussian precision. It leaves open an institutional alternative:
the regulator might obtain the relevant state from privately informed
producers, either through an authenticated certificate or through a
reporting scheme backed by random verification. We now make this choice
explicit. The same externality, the same ex post social objective, and
the same equilibrium amplitudes are used throughout. The added
institutional costs are real resource costs; fines and appropriately
financed reporting transfers are not treated as resources.

\subsection{Unverifiable messages and the implementation requirement}

At a given positive equilibrium amplitude $a$, full knowledge of the
branch $S\in\{-1,+1\}$ implies the optimal subsequent service
$t_S=ST(a)$, where $T(a)=\beta\sqrt a/\chi>0$. Suppose
$\nu>\omega\sqrt a$. The positive-population spokesperson strictly
prefers a higher signed service in both states, whereas the
negative-population spokesperson strictly prefers a lower one.
Messages are costless, unverifiable, and cannot be punished in the
basic communication institution.

\begin{proposition}[Unverifiable communication and implementation]
\label{prop:communication}
For a single spokesperson or a simultaneous pair of perfectly informed
spokespersons sending direct binary reports without contingent
transfers, a fully revealing truthful pure-strategy equilibrium is
impossible under the maintained opposing directional preferences.
If both spokespeople are instead paid $M$ when their reports agree,
a truth-telling equilibrium exists when
\begin{equation}
M>T(a)(\nu-\omega\sqrt a),\label{eq:agreebonus}
\end{equation}
with $t=0$ after mismatched reports and $t=ST(a)$ after matching
reports. Yet the agreement mechanism also admits an uninformative
pooling equilibrium with appropriate continuation beliefs.
\end{proposition}

We therefore evaluate information institutions by an explicit
\emph{reliable-implementation} requirement: reporting must be strictly
incentive compatible for a designated sender for every realized branch,
conditional on an enforceable precommitted verification policy.
Agreement payments alone do not meet this criterion, even if the
truthful equilibrium would cost no real resources. If the regulator
could freely coordinate on that equilibrium, it would replace all
costly information institutions whenever $a>0$. Robustness to
strategic reports consequently depends on the admissible mechanism
class, as the broader information-elicitation literature emphasizes
\citep{crawford_sobel1982,krishna_morgan2001,battaglini2002,miller2005}.

\subsection{Authenticated certification and verification-backed reports}

The confidential charge assessor observes an even activity-difference
statistic. It can release an authenticated signed report at a fixed
real cost $D_A\geq0$ only if given additional legal authority and
identification access. Alternatively, a designated positive-group
spokesperson may report the branch directly. The government commits
\emph{before} production choices to checking that report with a
probability $\rho$ independent of its content. A signed audit costs
$c_A>0$, detects an incorrect report, corrects the subsequent service
allocation, and imposes a collectible fine $F>0$ on the sender.
No fine is imposed on the truthful equilibrium path. Audits must be
enforceable commitments: without them, an uninformed regulator that
expects fully truthful reports would have no sequential incentive to
pay for an ex post check.

The maximum unilateral gain from falsely naming the favorable branch
at fixed $a$ is
\begin{equation}
G(a)=2T(a)(\nu-\omega\sqrt a)>0.
\label{eq:lying_gain}
\end{equation}
A false report is attractive only in the unfavorable state. It
raises the sender's payoff by $G(a)$ when no audit occurs, whereas a
signed audit corrects the allocation and collects the fine.
A fixed incentive margin $\delta\in(0,F)$ requires the truth to be
strictly preferable by at least $\delta$ in the unfavorable state.
This margin is a contractual implementation standard, measured in
money-metric private payoff, that insures the report against small
unmodelled gains from misrepresentation. Formally, if the true gain
from an undetected lie exceeds the modeled gain $G(a)$ by at most
$\varepsilon<\delta$, the prescribed audit probability still makes
a lie strictly unprofitable: its expected gain is at most
$-\delta+(1-\rho_\delta(a))\varepsilon<0$.
Thus a legal or administrative tolerance for bounded incentive
misspecification may determine $\delta$ before the regulator
optimizes enforcement. We hold this reliability requirement fixed
across institutions and policy choices. If the regulator were free
to demand only strict truth telling without any robustness standard,
it could approach $\delta=0$ from above, so its strict-IC program
would generally have an infimum rather than a strictly positive
cost-minimizing incentive margin. All substantive audit-regulation
comparisons below apply for each fixed $0<\delta<F$.

\begin{proposition}[Strict truthful reporting under committed audits]
\label{prop:auditIC}
Within the direct-report institution just described, strict truthful
reporting with incentive margin $\delta$ is implemented by the
minimum audit probability
\begin{equation}
\rho_\delta(a)=\frac{G(a)+\delta}{F+G(a)}\in(0,1).
\label{eq:strongaudit}
\end{equation}
The corresponding expected verification cost is
$c_A\rho_\delta(a)$. Conditional on a truthful report, the regulator
learns the branch perfectly and chooses $t_S=ST(a)$. This conclusion
requires a credible precommitment to the audit frequency, enforceable
and collectible fines, and access to a signed verification technology.
\end{proposition}

The guarantee applies to this one-sender direct mechanism; it does not
claim optimality over every feasible mechanism with several senders,
contingent payments, selective verification, or signal combinations.
Under the analytical baseline $\mu=0$, the producer coefficient
$\omega=0$ and $\beta=\zeta>0$; thus $G(a)=g\sqrt a$ with
$g=2\nu\beta/\chi>0$. This identity makes the resource cost of
credible reporting a known function of the equilibrium amplitude.

\subsection{The joint institutional and regulatory choice}

We solve the complete institutional-menu choice exactly for the
zero-private-anticipation benchmark $\mu=0$. Let $q>q_E$ and define
$E=E(q)>0$. The noninformational welfare component equals
\begin{equation}
\phi(a;q,\eta)=[-1-2\alpha(1-2q)-\eta]a-qa^2
-\frac\kappa2[E(q)-a]^2,
\quad 0\leq a\leq E.\label{eq:institution_phi}
\end{equation}
Because the regulatory charge and the selected information institution
are chosen jointly, each institution must optimize the activity
amplitude and hence the charge, rather than be evaluated at the
amplitude selected by another institution.

Let the values of optimal purchased Gaussian monitoring, full signed
certification before its fixed cost, and strict audit-backed reporting
be, respectively,
\begin{equation}
\begin{aligned}
V_G&=\max_{\substack{0\leq a\leq E\\\tau\geq0}}
\{\phi(a)+Ba f(a\tau)-C(\tau)\},\\
V_A^0&=\max_{0\leq a\leq E}\{\phi(a)+Ba\},\\
V_R(c_A)&=\sup_{0<a\leq E}
\{\phi(a)+Ba-c_A\rho_\delta(a)\}.
\end{aligned}\label{eq:institution_values}
\end{equation}
Full certification yields $V_C(D_A)=V_A^0-D_A$.
If the state has zero amplitude, no signed information is needed and
the Gaussian option with $\tau=0$ already supplies that outcome.
Consequently, the regulated economy's joint institution-and-policy
value is
\begin{equation}
V^*(q,\eta;D_A,c_A,F,\delta)
=\max\{V_G,\ V_A^0-D_A,\ V_R(c_A)\}.
\label{eq:envelope_institutions}
\end{equation}
This is the endogenous institutional choice within the specified
mutually exclusive menu. Random selection of an institution before
agents choose activities cannot improve on its maximum. Hybrid
schemes that use noisy public signals to screen private reports remain
outside the menu.

The perfectly informed benchmark is available in closed form. With
$D(q)=2q+\kappa$, $K(q)$ as in Section~\ref{sec:primitive}, and
$\Theta=K(q)-\eta+B$, define
$\widehat a=\min\{E,\max\{0,\Theta/D(q)\}\}$. Then
\begin{equation}
V_A^0=-\frac\kappa2 E^2+\Theta\widehat a
-\frac{D(q)}2\widehat a^2.
\label{eq:vagross}
\end{equation}

\begin{theorem}[Institutional choice and primitive cost frontiers]
\label{thm:institution_choice}
Assume $E(q)>0$, $\mu=0$, $F>\delta>0$, and $\nu>0$.
Define two real-cost thresholds
\begin{equation}
\begin{split}
D_{\mathrm{crit}}&=V_A^0-V_G\geq0,\\
c_{\mathrm{crit}}&=
\max_{0<a\leq E}
\frac{[\phi(a)+Ba-V_G]_+}{\rho_\delta(a)},
\quad [z]_+=\max\{z,0\}.
\end{split}\label{eq:institution_thresholds}
\end{equation}
Both thresholds are finite. Purchased Gaussian monitoring is
strictly preferable to certification when $D_A>D_{\mathrm{crit}}$
and certification is strictly preferable to it when
$D_A<D_{\mathrm{crit}}$.
Likewise, an active audit-backed disclosure institution strictly
dominates purchased monitoring when $c_A<c_{\mathrm{crit}}$;
all active audit choices are strictly inferior when
$c_A>c_{\mathrm{crit}}$.

If $V_A^0>V_G$, all three institutions are strictly optimal in
nonempty open regions of the verification-cost plane $(D_A,c_A)$.
At a nondegenerate network-induced Gaussian-policy crossing, if
\begin{equation}
D_A>D_{\mathrm{crit}}(q_0,\eta^*),\qquad
c_A>c_{\mathrm{crit}}(q_0,\eta^*),
\label{eq:institution_survival}
\end{equation}
then the original discrete regulation-and-monitoring switch remains
globally optimal within the expanded institutional menu for network
weights sufficiently close to $q_0$. If instead either alternative
strictly dominates Gaussian monitoring at the reference point,
that original crossing no longer describes the global joint policy
choice there.
\end{theorem}

The threshold $D_{\mathrm{crit}}$ prices a complete authenticated
record. The threshold $c_{\mathrm{crit}}$ prices the ability to enforce
a truthful private message through costly random inspections.
The latter threshold depends on the fine cap and the chosen strict
incentive margin, so it reflects the available legal and contractual
instruments rather than information technology alone.

\subsection{Audit incentives and optimal corrective regulation}

Two institutions that reveal the same signed state may nonetheless
induce different optimal activity choices. Signed certification has a
fixed verification cost, while truthful audit-backed reporting
requires a detection probability that rises with the private gains
from misrepresentation. This difference changes the marginal
social cost of allowing specialization.

\begin{proposition}[Audited reporting strengthens corrective regulation]
\label{prop:audit_regulation}
Maintain the zero-anticipation benchmark $\mu=0$, $E(q)>0$,
$0<\delta<F$, $g>0$, and $c_A>0$. Write
$h(a)=\phi(a)+Ba$, and suppose the certified-information
maximizer $a_C$ lies strictly inside $(0,E(q))$. If the
truthful-audit program admits an optimal positive amplitude
$a_R>0$, then
\begin{equation}
0<a_R<a_C, \qquad
r_R=E(q)-a_R>r_C=E(q)-a_C.
\label{eq:audit_regulation_order}
\end{equation}
The active audit optimum is interior and satisfies
\begin{equation}
D(q)(a_C-a_R)
=c_A\rho_\delta'(a_R)
=\frac{c_A(F-\delta)g}
{2\sqrt{a_R}\,[F+g\sqrt{a_R}]^2}>0,
\quad D(q)=2q+\kappa.
\label{eq:audit_regulation_wedge}
\end{equation}
Furthermore, for any two inspection costs
$0<c_A^1<c_A^2$ for which positive audit optima exist, every pair
of respective global maximizers satisfies
\begin{equation}
a_R(c_A^2)\leq a_R(c_A^1),\qquad
r_R(c_A^2)\geq r_R(c_A^1).
\label{eq:audit_regulation_coststat}
\end{equation}
The ordering in inspection costs does not require the audited
objective to be globally concave or its maximizer to be unique.
\end{proposition}

The result links information implementation to corrective taxation.
The certified and audited institutions produce the same ex post
public-service allocation, but under auditing the required enforcement expenditure rises
with specialization, creating an additional positive marginal cost. The regulator therefore implements a lower
activity difference under auditing and imposes a higher charge. As
inspection costs increase, the preferred active audit amplitude
may change discontinuously; the global monotonicity result remains
valid for any two active optima. The proposition does not rank
Gaussian monitoring against either full-information institution:
that comparison continues to depend on the resource-cost frontiers
in Theorem~\ref{thm:institution_choice}. It also does not apply
automatically if the certified optimum is on the boundary or the
audited program has no attained positive-amplitude optimum.

\begin{proposition}[Institutional substitution and enforcement]
\label{prop:institution_comparative}
Under the assumptions of Theorem~\ref{thm:institution_choice},
$V_R(c_A)$ is nonincreasing and convex in the per-inspection real
cost $c_A$, and is nondecreasing in the collectible fine cap $F$.
Moreover, as $c_A\downarrow0$ or $F\to\infty$,
\begin{equation}
V_R(c_A)\longrightarrow V_A^0
\quad\text{or}\quad
V_R(c_A;F)\longrightarrow V_A^0,
\label{eq:full_info_audit_limit}
\end{equation}
respectively. Provided $V_A^0>V_G$, sufficiently inexpensive
credible auditing replaces purchased Gaussian monitoring, whereas
sufficiently expensive auditing and certification preserve the
Gaussian-policy switch.

In addition, put
$\rho_{\min}=\delta/F$ and
$\rho_{\max}=(g\sqrt E+\delta)/(F+g\sqrt E)$.
If $D_A<c_A\rho_{\min}$, certification is strictly cheaper than
active audited reporting at every positive amplitude, so audited
reporting cannot be uniquely optimal. If
$D_A>c_A\rho_{\max}$, audited reporting is strictly cheaper than
certification at every positive amplitude, so certification cannot
be uniquely optimal.
\end{proposition}

At a regular interior optimum of the audit scheme, the envelope
derivative $\partial V_R/\partial c_A=-\rho_\delta(a_R)<0$.
The restriction to collectible fines is consequential: the formal
limit $F\to\infty$ is a theoretical benchmark rather than a claim
that sanctions are unbounded in an actual regulatory system.
These results connect the price of monitoring to the legal capacity
to verify and enforce truthful private information, rather than
maintaining the latter as an unexamined assumption.

\subsection{Forward-looking agents and institutional scope}

The preceding cost frontiers are exact for $\mu=0$. When $\mu>0$,
equilibrium amplitudes respond differently across institutions.
Under Gaussian monitoring the regulatory identity remains
$r=E(q)-a+(\mu/q)f(a\tau)$. Under truthful reporting or
full certification it becomes
\begin{equation}
r=E(q)-a+\mu/q,\label{eq:full_info_feedback}
\end{equation}
because producers anticipate perfect targeting of the realized
branch. Thus institutional choice must be recomputed with each
institution's own private-equilibrium constraint. Where the reference
policy optima are isolated and regular and both outside options have
strict value gaps, these gaps and the resulting institutional ranking
persist for sufficiently small positive $\mu$. The same conclusion
applies to the original network-induced policy switch under the
regularity assumptions of Theorem~\ref{thm:switch}. We make no
global primitive-threshold claim for arbitrary positive $\mu$.

Finally, the equilibrium-selection standard matters. If the
regulator could costlessly select the truthful equilibrium of the
agreement-payment mechanism, complete information would be acquired
without verification and would dominate purchased Gaussian precision
at a positive amplitude. Our institutional result therefore
characterizes public choices among reliably implemented technologies.
Removing enforceable audit commitments, allowing nonexclusive hybrid
information sources, or solving for all indirect mechanisms yields a
different design problem. These limitations are economically
substantive and delimit the scope of the paper.

\section{Policy Implications and Model Scope}

The corrective charge reduces the uninsured external risk generated by
specialization. Once that policy has determined the economic
configuration, information makes it possible to direct a public
service to the affected population. Under the Gaussian monitoring
institution, Theorem~\ref{thm:primitive} establishes primitive
conditions for a discontinuous switch between strong regulation
without purchased monitoring and more specialization with a purchased
signal. The comparative statics in Proposition~\ref{prop:signed}
identify how uninsured risk, information costs, and targeting benefits
shift the corresponding network exposure threshold.

The institutional comparison changes the scope of that result.
Theorem~\ref{thm:institution_choice} makes a signed disclosure
certificate and a strictly truthful audit-backed report endogenous
alternatives to Gaussian monitoring. Their expected cost includes
verification and enforceable reporting incentives, not simply the
technological price of gathering a noisy signal. Cheap certified
information or credible auditing can eliminate purchased monitoring.
If both alternatives remain expensive, however, the original
regulatory regime switch survives as an optimal policy even after
the institutional menu has been expanded. The regulator therefore
chooses the extent of corrective regulation and the organizational
means of obtaining state-contingent information jointly.

The institution-specific regulation comparison in
Proposition~\ref{prop:audit_regulation} adds a substantive
margin. When both certification and audited reporting reveal the
branch perfectly, the audit institution still calls for a stronger
corrective charge at an interior certified optimum, because its
probability of enforcement rises with specialization. Increasing the
cost of an inspection weakly raises that charge among active audit
optima, even if the audited objective is not globally concave.
This prediction does not require numerical calibration.

These comparisons rely on a deliberately restricted menu. The model
requires an enforceable ex ante audit commitment, an identifiable
reporter with a legally collectible finite fine, and a credible way to
verify the signed state when selected for audit. It evaluates fully
revealing reports that are strictly incentive compatible, while
agreement-contingent reporting rewards admit uninformative
alternative equilibria. Hybrid signal-and-report mechanisms, a fully
optimized indirect revelation mechanism, and direct insurance against
the households' uncontractible shocks fall outside the analysis.
The regulator's charge acts on a fixed network; the network links are
exogenous and the charge is an incentive instrument, not a direct
choice of network topology.

The private anticipation result is local and subject to the feedback
restrictions in Theorem~\ref{thm:anticipation}. Theorem~\ref{thm:primitive}
gives a primitive existence statement for the zero-feedback benchmark;
Theorem~\ref{thm:switch} establishes its conditional persistence under
small positive anticipation. When the regulator changes the
information institution, the private equilibrium must also be
recalculated because perfect and noisy information induce different
anticipated targeting returns. Section~\ref{sec:disclosure} states
the resulting constraint and limits its institutional comparison to
claims supported by the available regularity conditions.

\section{Conclusion}

A government choosing a corrective instrument also influences the
value of information that could be used after decentralized activity
has been determined. We model this relationship in a two-population
network economy where uninsurable risks borne by outside households
justify a charge on activity differences. Even when aggregate activity
obeys an exact stable law, the two specialization equilibria require
opposite state-contingent public-service decisions. Under explicit
primitive conditions, jointly chosen regulation and Gaussian
monitoring exhibit a discontinuous switch between optimal policy
regimes. The result persists locally under appropriately bounded
anticipation of future policy.

The institutional extension identifies a second margin of the
regulatory problem. The state may buy a noisy signal, obtain a
certified signed state, or elicit that state through a strictly
incentive-compatible reporting mechanism backed by committed random
audits. The resource-cost thresholds determine when certification or
auditing displaces purchased monitoring, and the original Gaussian
policy switch is preserved when both verification alternatives are
sufficiently expensive. More importantly, an audited report and a
certificate provide the same complete information but induce
different ex ante policy choices: when the certified optimum is
interior and the audited program actively operates, auditing calls
for less specialization and a higher corrective charge. Higher
inspection costs reinforce regulation among active audited choices.
The availability and enforceability of an information institution
therefore change both the attainable information and the optimal
corrective response.

The result is conditional on the block-symmetric network,
producer-private selection of the equilibrium branch, a specific
information institution menu, and the ability to commit to a
random-audit policy. The paper does not solve a general mechanism
design problem, select among multiple cheap-talk equilibria, or
establish global comparative statics for arbitrarily strong private
anticipation. The economic implication is therefore precise:
corrective charges, the information needed for contingent public
services, and the institution that supplies that information must be
assessed jointly. Exact aggregation alone cannot resolve this
public-finance decision.

\section*{Acknowledgments}
The author received no funding except from the organizations with which he is affiliated.

\section*{Declarations}
\noindent\textbf{Competing interests.} The author declares no conflicts of interest relevant to this manuscript.

\noindent\textbf{Data availability.} No empirical data were used or generated. This study is theoretical, and no external dataset is required to reproduce its analytical results.

\noindent\textbf{Use of artificial intelligence.} During the preparation of this work the author used ChatGPT (OpenAI) to review theoretical derivations, refine manuscript structure, and improve readability. The author independently verified the derivations and reviewed and edited the AI-assisted output as needed, and takes full responsibility for the content of the published article.

\appendix

\section{Proofs for Communication and Institution Choice}\label{app:disclosure}

\paragraph{Proposition~\ref{prop:communication}.}
Write $t_{ij}$ for the government's service after binary messages
$i,j\in\{+,-\}$ from the two populations. Under complete truthful
revelation, $t_{++}=T(a)>0$ and $t_{--}=-T(a)<0$.
In state $S=+$ the negative group must not gain by changing its
message to $-$, which requires $t_{++}\leq t_{+-}$. In state $S=-$,
the positive group must not gain by changing its message to $+$,
which requires $t_{--}\geq t_{+-}$. The inequalities contradict
$t_{++}>t_{--}$. The same monotone directional preference precludes
fully revealing direct truthful reports from a single sender without
state-contingent monetary incentives.

For agreement rewards, set $t_{++}=T(a)$, $t_{--}=-T(a)$, and
$t_{+-}=t_{-+}=0$, paying $M$ to each sender following agreement.
A sender who disfavors the truthful fully informed service gains at
most $T(a)(\nu-\omega\sqrt a)$ by deviating to a disagreement, but
loses the reward $M$. At the stated strict inequality deviation is
unprofitable. Favorable-state deviation is unprofitable even without
the reward. Hence there exists a truthful equilibrium.
A pooling strategy in which both senders report $+$ for every branch
can also be an equilibrium: choose continuation allocation zero
after the pooling report and after mismatches, with compatible
off-path beliefs. A unilateral deviation then changes neither the
service nor the truth of the branch but loses $M$. The informative
equilibrium is therefore not uniquely implemented.\qed

\paragraph{Proposition~\ref{prop:auditIC}.}
Given true unfavorable state $S=-$, a false favorable report changes
the service from $-T(a)$ to $+T(a)$ if no audit occurs. Its private
gain is $G(a)$. If the regulator audits, it learns the true branch,
corrects the action, and imposes fine $F$. Relative to truth telling,
the expected gain from lying is
$(1-\rho)G(a)-\rho F=G(a)-\rho(F+G(a))$.
For this difference to be at most $-\delta$, the audit probability
must satisfy $\rho\geq(G(a)+\delta)/(F+G(a))$.
The minimum lies strictly between zero and one since
$0<\delta<F$. In the favorable branch the sender strictly prefers
the truthful higher service and also avoids the possible fine, so
truth is uniquely optimal there as well. Under the precommitted
verification rule this yields strictly truthful reporting in both
branches and full identification of $S$ on the equilibrium path.
The real expected verification cost is $c_A\rho_\delta(a)$, while
fines are zero on that path. The conclusion does not follow without
the stated commitment and enforceability.\qed

\paragraph{Theorem~\ref{thm:institution_choice}.}
With $\mu=0$, the regulatory identity is $r=E(q)-a$, so the same
baseline welfare $\phi(a;q,\eta)$ is appropriate for all three
information institutions. Gaussian monitoring gives posterior-based
expected targeting gain $Ba f(a\tau)$, minus its purchase cost.
Certification reveals the sign and delivers gross gain $Ba$ with
fixed resource cost $D_A$. An incentive-compatible audited report
also reveals the sign but requires the expected resource cost
$c_A\rho_\delta(a)$ at the cheapest audit probability meeting the
chosen strict incentive margin. Optimizing $a$ and, under Gaussian
monitoring, $\tau$, gives equation~\eqref{eq:envelope_institutions}.

The Gaussian value is at most $V_A^0$ because $f\leq1$ and
monitoring has nonnegative cost. Both are finite because
$a\in[0,E]$, Gaussian purchases have coercive quadratic cost, and
$\phi$ is continuous. Full certification beats Gaussian monitoring
exactly if $D_A<V_A^0-V_G$.
For audited reporting, the payoff at a positive amplitude exceeds
$V_G$ if and only if
$c_A<[\phi(a)+Ba-V_G]/\rho_\delta(a)$.
Maximizing the nonnegative numerator divided by positive
$\rho_\delta(a)$ therefore gives the stated exact threshold.
Since $\rho_\delta(a)\geq\delta/F>0$, the ratio is bounded and the
threshold is finite. A cost strictly above the threshold makes every
active audit choice strictly inferior; a cost strictly below produces
an active audit choice with strictly higher welfare.

When $V_A^0>V_G$, positive certification and auditing cost thresholds
exist. With both costs high Gaussian monitoring is strictly optimal
among active institutions. With certification cost near zero and
audit cost high, certification is strictly optimal. With certification
cost high and inspection cost near zero, audited reporting is
strictly optimal because $V_R(c_A)\to V_A^0>V_G$.
These strict comparisons persist over nonempty open sets of
institutional costs. If both alternatives are strictly worse at a
regular Gaussian switching point, continuity of the three optimized
values in network exposure (on the relevant compact feasible
neighborhood) preserves both strict gaps. The low- and high-amplitude
Gaussian choices therefore remain globally optimal within the menu
and continue to exchange their ranking at the original network
crossing. If an alternative is strictly better, Gaussian optimality
at that point fails by definition.\qed

\paragraph{Proposition~\ref{prop:audit_regulation}.}
At zero private anticipation the optimal regulatory identity is
$r=E(q)-a$. Hence both complete-information technologies share the
gross welfare function
$h(a)=\phi(a)+Ba=-\kappa E(q)^2/2+\Theta a-D(q)a^2/2$.
Its unique unconstrained maximum is at $a_C=\Theta/D(q)$, which
is interior by hypothesis, and $h''(a)=-D(q)<0$.
The minimum reliable audit probability is
$\rho_\delta(a)=(g\sqrt a+\delta)/(F+g\sqrt a)$, with
\[
\rho_\delta'(a)=
\frac{(F-\delta)g}{2\sqrt a\,(F+g\sqrt a)^2}>0
\quad\text{for all }a>0.
\]
For $a>a_C$, the gross welfare $h(a)$ is strictly smaller than
$h(a_C)$, while the verification cost is strictly higher. Thus an
audited global optimum cannot lie above $a_C$. At $a_C$, the right-
and left-derivatives of $h(a)-c_A\rho_\delta(a)$ are negative,
so $a_C$ itself cannot maximize this objective, given interiority
of $a_C$. By hypothesis a positive audited optimum is attained;
it follows that $0<a_R<a_C$, and therefore $r_R>r_C$.
The audited optimum is interior and its first-order condition is
$h'(a_R)=c_A\rho_\delta'(a_R)$. Substituting
$h'(a)=D(q)(a_C-a)$ yields equation~\eqref{eq:audit_regulation_wedge}.

For the global cost comparison let $a_1$ be any optimal positive
audit amplitude when the cost is $c_A^1$ and $a_2$ any optimal
positive audit amplitude when the cost is $c_A^2>c_A^1$.
Optimality gives
\[
\begin{aligned}
h(a_1)-c_A^1\rho_\delta(a_1)
 &\geq h(a_2)-c_A^1\rho_\delta(a_2),\\
h(a_2)-c_A^2\rho_\delta(a_2)
 &\geq h(a_1)-c_A^2\rho_\delta(a_1).
\end{aligned}
\]
Adding the inequalities yields
$(c_A^2-c_A^1)[\rho_\delta(a_1)-\rho_\delta(a_2)]\geq0$.
Because $\rho_\delta$ is strictly increasing,
$a_2\leq a_1$, proving the weak monotonicity of specialization
and the opposite weak monotonicity of regulation. No second-order
condition or uniqueness of the audited optimum is used.\qed

\paragraph{Proposition~\ref{prop:institution_comparative}.}
For every amplitude $a>0$, the audited value is affine and decreasing
in $c_A$, with slope $-\rho_\delta(a)<0$. The supremum of these
functions is therefore nonincreasing and convex in $c_A$. Because
$\partial\rho_\delta/\partial F
=-(G(a)+\delta)/(F+G(a))^2<0$, it is nondecreasing in the fine cap.
With finite $E$, the lie gain $G(a)$ is bounded, so $\rho_\delta(a)$
converges to zero uniformly as $F\to\infty$. It also follows that
$c_A\rho_\delta(a)\to0$ uniformly as $c_A\downarrow0$.
The optimized audited value consequently approaches $V_A^0$
in either limit.

At $\mu=0$, $G(a)=g\sqrt a$ and $\rho_\delta(a)$ is increasing in
$a$, since its derivative with respect to $G$ is
$(F-\delta)/(F+G)^2>0$. Therefore all positive-amplitude audited
verification costs lie between $c_A\delta/F$ and
$c_A(g\sqrt E+\delta)/(F+g\sqrt E)$.
When $D_A$ is strictly smaller than the lower bound, certification
is strictly better for each common positive amplitude. Conversely,
when $D_A$ strictly exceeds the upper bound, audit-backed reporting
is strictly better at each common positive amplitude. Since Gaussian
monitoring already includes the uninformed zero-amplitude allocation,
none of these pointwise comparisons implies that an institution
provides strictly positive welfare gains in every parameter region.
The assertions concern institutional selection within the menu.\qed

\section{Proof of the Primitive Switch and Threshold Effects}
\label{app:primitive}

\paragraph{Entry threshold and local loss of concavity.}
The Gaussian estimation identity yields $f(v)\leq v$ and $f'(0)=1$.
It follows that, for $a\leq a_0$,
$Ba f(a\tau)-C(\tau)\leq(Ba^2-c_1)\tau-c_2\tau^2/2\leq0$,
so $\Psi(a)=0$. For $a>a_0$, its derivative in $\tau$ at zero is
$Ba^2-c_1>0$, so $\Psi(a)>0$. Expanding $f(v)=v-v^2+O(v^3)$,
we obtain, as $a$ approaches $a_0$ from above,
\begin{equation}
\Psi(a)=\frac{(Ba^2-c_1)^2}{2(2Ba^3+c_2)}
+o((a-a_0)^2),\qquad
\Psi''(a_0+)=\frac{4B^2a_0^2}{2Ba_0^3+c_2}.
\label{eq:psilocalproof}
\end{equation}
At $a=a_0$, $\Psi$ is continuously differentiable with
$\Psi'(a_0)=0$. Under~\eqref{eq:primitive_conditions},
$\overline{\W}''(a_0+)>0$. Hence $a_0$ cannot be a maximum
for any linear-welfare coefficient $\theta=K(q_0)-\eta$:
when $\theta\neq D(q_0)a_0$ its first derivative is nonzero,
and when equality holds its right-hand curvature is positive.

\paragraph{Existence at positive risk costs.}
Fix $q=q_0$, abbreviate $D=D(q_0)$ and $E=E(q_0)$, and omit
the common constant in~\eqref{eq:wprofileprimitive}. Define the
restricted value functions
\begin{equation}
V_L(\theta)=\max_{0\leq a\leq a_0}
\left\{\theta a-\tfrac D2a^2\right\},\qquad
V_H(\theta)=\max_{a_0\leq a\leq E}
\left\{\theta a-\tfrac D2a^2+\Psi(a)\right\}.
\end{equation}
They are finite and continuous. For sufficiently negative $\theta$,
$V_L(\theta)>V_H(\theta)$; for sufficiently positive $\theta$,
$V_H(\theta)>V_L(\theta)$. By continuity, some $\theta^*$
equalizes them. This crossing is unique: the high-class maximizer
never has a smaller amplitude than the low-class maximizer, so the
high-minus-low value is nondecreasing in $\theta$; a zero interval
would require $a_0$ to be a global optimum, which was excluded. As $a_0$ is never a maximum of the full objective,
the maximizing amplitudes at this crossing satisfy
$0\leq a_L<a_0<a_H\leq E$, and information acquisition is
strictly positive at $a_H$. A crossing with $\theta^*\geq Da_0$
would force the restricted low-information maximum to equal $a_0$,
contradicting the strict separation. Hence $\theta^*<Da_0$.
Since $K(q_0)>Da_0$, the implied risk parameter
$\eta^*=K(q_0)-\theta^*$ is strictly positive.

\paragraph{Network-induced switching.}
At $\mu=0$ the feasible interval $[0,E(q)]$ expands strictly with
$q$, because $E'(q)=(\alpha+1)/q^2>0$. On this interval
\eqref{eq:increasingdifferences} proves strictly increasing differences
of the profiled objective in $(q,a)$. For $q>q_0$ sufficiently close,
$a_H$ remains feasible and its objective exceeds that of every
$a\leq a_0$ by strict increasing differences. For $q<q_0$ sufficiently
close, $a_L$ remains feasible and dominates every feasible
$a\geq a_0$ by the same comparison. Therefore every global optimizer
lies strictly below $a_0$ on one side of $q_0$ and strictly above
$a_0$ on the other. Since $a_0$ is never an optimum, the transition
is discontinuous.

\paragraph{Persistence under positive anticipation.}
At $q_0$ the closed interval around $a_0$ contains no global optimizer;
compactness yields a strictly positive welfare gap over a smaller
neighborhood of this amplitude. Coercive quadratic monitoring costs
allow the feasible signal precision to be restricted uniformly to a
finite bound for parameters near $(q_0,0)$. The feasible policy
correspondence and welfare are continuous in $q$ and $\mu$ as
$\mu\downarrow0$. Choose network weights $q_-<q_0<q_+$ close
enough that at $\mu=0$ all global optima at $q_-$ are below $a_0$
and all at $q_+$ are above $a_0$. These strict separations and the
central welfare gap persist for sufficiently small positive $\mu$.
The values of the low- and high-amplitude restricted optimization
problems are continuous in $q$, and their ranking reverses between
$q_-$ and $q_+$. Their equality at some intermediate $q^*(\mu)$
is therefore a global two-regime switch. The exact aggregate
equation follows from Theorem~\ref{thm:aggregation}.

\paragraph{Comparative statics.}
At an interior, nondegenerate two-regime crossing, write
$\Delta V=V_H-V_L$. The envelope theorem and
\eqref{eq:wprofileprimitive} yield
\begin{equation}
\partial_q\Delta V=(a_H-a_L)M>0,
\qquad \partial_\eta\Delta V=-(a_H-a_L)<0.
\end{equation}
As the low-information optimum has $\tau_L=0$, the envelope
derivatives with respect to information costs are
$\partial_{c_1}\Delta V=-\tau_H$ and
$\partial_{c_2}\Delta V=-\tau_H^2/2$.
Similarly $\partial_B\Delta V=a_Hf(a_H\tau_H)$.
From the reduced welfare~\eqref{eq:reduced}, at $\mu=0$,
$\partial_\mu\Delta V=-\kappa r_H f(a_H\tau_H)/q_0$.
Implicit differentiation of $\Delta V(q^*,\cdot)=0$ gives all
formulas in Proposition~\ref{prop:signed}. The strict signs continue
locally by continuity.\qed

\section{Proof of Proposition~\ref{prop:baseline}}
For two independent Gaussian signals $Y_j=d+\xi_j$, Bayes' rule and the independent likelihoods give a combined log-likelihood ratio equal to $2\sqrt a(\tau_0Y_0+\tau Y_1)$. The posterior mean therefore has the form in~\eqref{eq:free_posterior}. Under either branch, the combined normalized score has mean $a(\tau_0+\tau)$ and variance $a(\tau_0+\tau)$, so the ex ante posterior square is $a f(a(\tau_0+\tau))$. The conditional-variance decomposition gives equation~\eqref{eq:baseline_increment}.

At $\tau_0=\mu=0$, the low-precision maximizer has $a_L<a_0$, hence its one-sided acquisition derivative equals $Ba_L^2-c_1<0$. The high-precision maximizer has an interior, nonsingular negative-definite Hessian. The objective and equilibrium constraint depend continuously and smoothly on $(q,\tau_0,\mu)$ in neighborhoods of the two optima (with the one-sided derivative understood at $\tau=0$). The implicit-function theorem continues the two local maximizers; the low remains on the boundary $\tau=0$, and the high retains $\tau>0$.

Monitoring costs are coercive, and the feasible amplitude remains bounded in a sufficiently small neighborhood of the benchmark. Thus all global optimizers are confined to a common compact set. Strict separation between the reference optima and other feasible alternatives implies persistence of the two candidates as the only global candidates near $q_0$. The derivative of their value difference with respect to $q$ is nonzero by the transverse-crossing assumption. Another application of the implicit-function theorem gives the nearby switching weight, with a reversal of value ranking across it. For $\mu>0$ sufficiently small, the regular equilibrium branch and the corresponding charge remain feasible by the same continuity argument. This establishes the proposition.\qed

\section{Proof of Theorem 1}

Because the interaction operator is symmetric with row sums one, its column sums also equal one. Let a dot denote the gradient adjustment to expected private marginal utility. Summing private marginal returns over the continuum, the direct quadratic term gives $-m$ and the strategic term gives $-\alpha m$. For every symmetric pair $(i,j)$, the cubic difference term for $i$ is the negative of the term for $j$; the same is true of the linear difference charge. Their aggregate integrals therefore vanish. The future targeted term contributes $\omega\E[t\mid S]$ to agents in one equally sized group and $-\omega\E[t\mid S]$ to agents in the other. It also cancels. This establishes \eqref{eq:agg} for arbitrary profiles.

Strict own-action concavity follows from \eqref{eq:concavity}. Uniform interactions within each population imply common strictly concave best-response problems, so population-homogeneous equilibrium is necessary. At a profile $x_+=m+d$, $x_-=m-d$, the other-agent mean entering the payoff of a $+$ agent is $m+(1-2q)d$, and that entering a $-$ agent is $m-(1-2q)d$. The difference-charge contribution to the $+$ agent's marginal return is $-rqd$; the quartic contribution is $-qd^3$. Taking the mean of the two marginal-return equations gives $-(1+\alpha)m$. Taking their half-difference gives
\begin{equation}
\dot d=[\varepsilon(q)-qr]d-qd^3
+\omega\E[t\mid S].
\label{eq:ddyn}
\end{equation}
For $d=S\sqrt a$ and \eqref{eq:post}, the last term equals $\mu S\sqrt a f(a\tau)$, yielding \eqref{eq:amp}.

For $\mu=0$ with $\varepsilon(q)>qr$, equation \eqref{eq:ddyn} has stationary values $d=0$ and $d=\pm\sqrt{[\varepsilon(q)-qr]/q}$. Its derivative with respect to $d$ is $\varepsilon(q)-qr-3qd^2$, which is positive at zero and negative at the two nonzero states. The mean direction has eigenvalue $-(1+\alpha)<0$. A zero-mean within-group perturbation, holding continuation beliefs fixed, has the strictly negative own-action linearization $-1-r/2-3qa/2$. Thus each asymmetric state is locally stable under the stated adjustment.\par
Conditional on the exogenous producer-private coordination signal $S$, the continuation expectations are held fixed while the gradient adjustment takes place. The scalar cubic then has one attracting solution on the half-line $Sd>0$; at $d=0$ its field has the sign of $S$, and for sufficiently large $|d|$ it points toward the origin. This establishes local attraction to the selected branch within the stipulated adjustment, but the equal prior probabilities come from the fair sunspot rather than a claim that the adaptive dynamics endogenously choose the branch.\qed

\section{Proof of the Bayesian Identity and Theorem 2}

Let $S$ be uniform on $\{-1,1\}$ and define a normalized observation $Z=vS+\sqrt vN$. Bayes' rule gives $\E[S\mid Z=z]=\tanh z$. Therefore
\begin{equation}
\E[S\tanh Z]=\E[(\E[S\mid Z])^2]=\E[\tanh^2 Z]=f(v).
\end{equation}
Symmetry implies $\E[\tanh Z\mid S]=Sf(v)$, proving \eqref{eq:post} after the change of variables $v=a\tau$. Equivalently, $f(v)=\E_N[\tanh(v+\sqrt vN)]$.

For $v>0$, differentiate the last formula under the integral, then apply Gaussian integration by parts:
\begin{equation}
\begin{split}
f'(v)
&=\E_N\left[\operatorname{sech}^2 Z\left(1+\frac{N}{2\sqrt v}\right)\right]\\
&=\E_N[\operatorname{sech}^2 Z(1-\tanh Z)]
=\E_N[\operatorname{sech}^4 Z],
\end{split}
\label{eq:fprime}
\end{equation}
where the final equality uses Bayes' rule to relate the conditional and unconditional expectations of even functions of $Z$. Thus $0<f'(v)\leq1$. The identity extends continuously to $v=0$, where $f'(0)=1$.

Write $F(a,r,\tau)=qa-\mu f(a\tau)-\varepsilon(q)+qr$. Under $q>\mu\tau$, its derivative $F_a=q-\mu\tau f'(a\tau)>0$ for every $a\geq0$. Since $F(0)=-(\varepsilon(q)-qr)<0$ and $F(a)\to\infty$ as $a\to\infty$, there is one and only one positive root. The implicit-function theorem yields
\begin{equation}
F_a\frac{\partial a}{\partial\tau}-\mu a f'(a\tau)=0,
\qquad
F_a\frac{\partial a}{\partial r}+q=0,
\end{equation}
which prove \eqref{eq:dat} and \eqref{eq:dar}. Holding the equilibrium continuation belief fixed under short-run gradient adjustment, the derivative of \eqref{eq:ddyn} at $d=S\sqrt a$ is
\begin{equation}
\varepsilon(q)-qr-3qa=-2qa-\mu f(a\tau)<0.
\end{equation}
The mean and within-group directions are strictly stable as shown in Appendix A. The claimed local stability and comparative statics follow. \qed

\section{Derivation of the Welfare Function}
\paragraph{Sign-invariant tax assessment and information reports.}
At a block-constant profile the squared-difference tax base of either population is
$\int w_{ij}(x_i-x_j)^2\,dj=4qa$. It therefore identifies the absolute
amplitude while revealing no sign of $d$. The group-level distributional
payoffs from a differential service are $+\nu t$ and $-\nu t$, so they
cancel in the two-group welfare sum and have zero derivative with respect to
private activity. If $\nu>\omega\sqrt{a_{\max}}$, the private slope of
continuation payoff in $t$ is positive for every agent in the first group
and negative for every agent in the second, regardless of the realized
branch. Thus a fully revealing single-sender report of the branch cannot
be incentive compatible: each group has a state-independent preference
for one direction of service allocation. The policymaker is assumed to
observe only the certified even tax base, not the labeled underlying activities.

At a block-constant profile with $m=0$ and $x_+=d$, $x_-=-d$, each representative group's individual base payoff is
\begin{equation}
-\tfrac12d^2-\alpha(1-2q)d^2-\tfrac q2d^4-rqd^2.
\end{equation}
Summing across the two groups gives equation \eqref{eq:privatew}. Rebate revenue equals $2rqa$ and cancels the private payment. The two CARA households' certainty-equivalent losses sum to $\eta a$ by \eqref{eq:eta}. Ex post net surplus conditional on $Y$ is the quadratic function
\begin{equation}
\frac\beta\chi\E[d\mid Y]\,\beta\E[d\mid Y]
-\frac\chi2\left(\frac\beta\chi\E[d\mid Y]\right)^2
=\frac{\beta^2}{2\chi}\E[d\mid Y]^2.
\end{equation}
Its ex ante mean is $Ba f(a\tau)$ by \eqref{eq:post}. Adding administrative costs $\kappa r^2/2$ and information costs $C(\tau)$ gives \eqref{eq:welfare}. Substituting \eqref{eq:rfroma} gives \eqref{eq:reduced}. \qed

\section{Proof of Theorem~\ref{thm:switch}}

At $\mu=0$, the equality $r=E(q)-a$ restricts the amplitude to $0\leq a\leq E(q)$ and converts \eqref{eq:reduced} into
\begin{equation}
\calF_0(a,\tau;q)
=\bigl[-1-2\alpha(1-2q)-\eta\bigr]a-qa^2
-\tfrac\kappa2[E(q)-a]^2
+Ba f(a\tau)-C(\tau).
\label{eq:fzero}
\end{equation}
For the low-information optimizer $z_L=(a_L,0)$, the strict one-sided derivative and strict curvature in $a$ imply a locally unique maximizer that remains on the boundary $\tau=0$ for nearby parameter values. For the interior optimizer $z_H$, the negative definite Hessian permits application of the implicit-function theorem to the two first-order conditions. These give smooth local solution branches $z_j(q,\mu)$, $j=L,H$, for parameters near $(q_0,0)$. The structural charge remains positive by continuity, as does the weak-feedback regularity condition for sufficiently small $\mu$.

For $\mu=0$, partial differentiation of \eqref{eq:fzero} at a fixed $(a,\tau)$ yields
\begin{equation}
\partial_q\calF_0=4\alpha a-a^2
-\kappa[E(q)-a]\frac{\alpha+1}{q^2}.
\label{eq:qder}
\end{equation}
The envelope theorem applies to each of the two local optimum-value branches. Their difference $\Delta(q,\mu)=V_H(q,\mu)-V_L(q,\mu)$ satisfies $\Delta(q_0,0)=0$, and subtracting \eqref{eq:qder} evaluated at the two optimizers gives
\begin{equation}
\partial_q\Delta(q_0,0)
=(a_H-a_L)\left[4\alpha-(a_H+a_L)
+\kappa\frac{\alpha+1}{q_0^2}\right]=T_0\ne0.
\end{equation}
The implicit-function theorem therefore delivers a unique local switching curve $q^*(\mu)$ satisfying $\Delta(q^*(\mu),\mu)=0$. Its crossing is transverse, so the preferred branch reverses locally.

It remains to justify global rather than merely local optimality. In the reference model, the two isolated global optima are assumed to be separated from all choices outside their disjoint neighborhoods by a strict welfare gap. In the nearby models, $a$ stays bounded because $r\geq0$ implies $a\leq E(q)+\mu/q$; the quadratic information cost makes a common finite upper bound on relevant $\tau$ possible. The objective and feasible constraints vary continuously in the parameters. Uniform continuity on this compact, effectively feasible set preserves the welfare gap for nearby parameters. Hence the two continued local optima exhaust the global candidates, and the locally crossing values are globally optimal at $q^*(\mu)$. Exact aggregation is unaffected by Theorem~\ref{thm:aggregation}. \qed

\section{Proof of Proposition~\ref{prop:gaussian}}

The posterior identity in Appendix B gives $f(0)=0$, $f'(v)\leq1$, and therefore $f(v)\leq v$. For every $a\leq a_0=\sqrt{c_1/B}$ and $\tau\geq0$,
\begin{equation}
Ba f(a\tau)-c_1\tau-\tfrac{c_2}{2}\tau^2
\leq (Ba^2-c_1)\tau-\tfrac{c_2}{2}\tau^2\leq0.
\end{equation}
Hence $\Psi(a)=0$ on that interval. For $a>a_0$, the derivative of the information surplus at $\tau=0$ is $Ba^2-c_1>0$, so $\Psi(a)>0$.

Expand the posterior quality around zero. Differentiating the Gaussian expectation or expanding $\tanh$ under the normal integral gives $f(v)=v-v^2+O(v^3)$. Thus
\begin{equation}
Ba f(a\tau)-C(\tau)
=(Ba^2-c_1)\tau
-\tfrac12(2Ba^3+c_2)\tau^2+O(\tau^3).
\end{equation}
The maximizing $\tau$ for $a\downarrow a_0$ from above is
\begin{equation}
\tau^*(a)=\frac{Ba^2-c_1}{2Ba^3+c_2}+O((a-a_0)^2),
\end{equation}
so
\begin{equation}
\Psi(a)=\frac{(Ba^2-c_1)^2}{2(2Ba^3+c_2)}+o((a-a_0)^2).
\end{equation}
Differentiating the leading expression twice yields \eqref{eq:curv}. At $\mu=0$, the part of welfare not involving $\Psi$ has second derivative $-(2q+\kappa)$, proving the nonconvexity criterion \eqref{eq:nonconvex}. It is an implication about the geometry of the policy objective, not a guarantee of a specific global-switch location. \qed

\section{Analytical Profiling of the Regulatory Problem}
\label{app:profiling}

The two-dimensional policy problem admits a useful exact reduction.
For $a>0$, set $v=a\tau$, the signal-to-noise ratio of the public
observation. Let $L(q)=-1-2\alpha(1-2q)-\eta$. Substituting
$\tau=v/a$ into~\eqref{eq:reduced} yields
\begin{equation}
\begin{split}
\widetilde{\calF}(a,v;q,\mu)={}&L(q)a-qa^2
-\frac\kappa2\left[E(q)-a+\frac\mu qf(v)\right]^2\\
&+Ba f(v)-c_1\frac va-\frac{c_2}{2}\frac{v^2}{a^2}.
\end{split}
\label{eq:profile_analytic}
\end{equation}
For each fixed $v\geq0$ and any admissible $a>0$,
\begin{equation}
\frac{\partial^2\widetilde{\calF}}{\partial a^2}
=-(2q+\kappa)-\frac{2c_1v}{a^3}
-\frac{3c_2v^2}{a^4}<0.
\label{eq:profile_concavity_analytic}
\end{equation}
For a fixed $v$, the charge constraint is the interval restriction
$0<a\leq E(q)+\mu f(v)/q$. Hence the inner maximization over
$a$ is strictly concave. This observation does not imply concavity
of the remaining profiled problem in $v$, and the discontinuity result
follows from the primitive inequalities and the proof of
Theorem~\ref{thm:primitive}, rather than from this change of variables.
When $a=0$, the best information choice is $\tau=0$, which is
covered separately by the no-monitoring benchmark.

\begingroup
\setstretch{1.03}

\endgroup
\end{document}